\documentclass{ws-procs11x85}
\usepackage{balance}

\makeindex

\newcommand{\met}{\hbox{E\kern-0.5em\lower-0.1ex\hbox{/}}_T}

\begin{document}

\title{THE BAYESIAN EFFECTS IN MEASUREMENT OF THE ASYMMETRY OF POISSON FLOWS}

\author{S. I. Bityukov \and V. V. Smirnova}

\address{Institute for high energy physics, 142281 Protvino, Russia\\
E-mail: Serguei.Bitioukov@cern.ch, vera@cub.ihep.su}

\author{N. V. Krasnikov}

\address{Institute for nuclear research RAS, 117312 Moscow, Russia\\
E-mail: krasniko@ms2.inr.ac.ru}

\author{A. Kuznetsov}

\address{Dept.of Physics, Northeastern University, Boston, MA 02115, USA}


\twocolumn[\maketitle\begin{abstract} 
As it follows from the interrelation of Gamma and Poisson 
distributions~~\cite{Bit3,1} the observed value of asymmetry for 
Poisson flows of events has the bias. The Monte Carlo experiment 
confirms the presence of this bias between the observed 
and initial asymmetries.
The obtained results allow to 
correct the estimation of the asymmetry
of flow rates.
\end{abstract}]

\bodymatter

\baselineskip=13.07pt
\section{Introduction}\label{sec:intr}

In the report the usage of the properties of the statistically dual 
distributions~\cite{Bit1}
(Poisson and Gamma) as well as the concept the ``confidence density of 
parameter''~\cite{Efr} allows us to show the presence of a bias in 
reconstruction of the initial asymmetry, which produced the 
observed asymmetry. We use here the confidence density as a posteriori
density with assumption that we have uniform prior.

\section{The modeling of the asymmetry}\label{sec:model}

Under the {\it initial asymmetry} we keep in mind the difference between the 
relative mean numbers of events 
($\displaystyle \frac{\mu_1}{\mu_1+\mu_2}$ and 
$\displaystyle \frac{\mu_2}{\mu_1+\mu_2}$) of two different flows of 
events: 

\begin{equation}
A = \frac{\mu_1-\mu_2}{\mu_1+\mu_2}. 
\label{eq:1}
\end{equation}

Under the {\it observed asymmetry} we keep in mind the difference between the 
relative {\it observed} numbers of events 
($\displaystyle \frac{n_1}{n_1+n_2}$ and 
$\displaystyle \frac{n_2}{n_1+n_2}$) from the same pair of flows 
of events: 

\begin{equation}
\hat A = \frac{n_1-n_2}{n_1+n_2}.
\label{eq:2}
\end{equation}

In ref~\cite{Bit1} (see, also,~\cite{1,2}) it is shown that the confidence 
density of the Poisson distributed parameter $\mu$ in the case of a single
observation $n$ is the Gamma distribution  $\Gamma_{1,1+n}$ with mean, mode, 
and variance $n+1,~n$, and $n+1$ respectively. 
This statement was checked by the Monte Carlo experiment~\cite{ACAT2}.

The difference between the most probable and mean values of 
the parameter of the Poisson distribution
suggests that there takes place 
(in the case of the measurement of asymmetry) a deviation 
which can be approximately estimated by the expression

\begin{equation}
\hat A_{cor} = \frac{(n_1+1)-(n_2+1)}{(n_1+1)+(n_2+1)} = 
\hat A \cdot \frac{n_1+n_2}{n_1+n_2+2},
\label{eq:3}
\end{equation}

\noindent
where values $n_1$ and $n_2$ are the observed numbers
of events from two Poisson distributions with parameters $\mu_1$
and $\mu_2$ correspondingly.

We carried out the uniform scanning  of parameter $A$, varying $A$ from 
value $A=-1$ to value $A=1$ using step size 0.01.  
By playing with the two Poisson distributions 
(with parameters $\mu_1$ and $\mu_2$) and using 30000 trials 
for each value of $A$ we used the RNPSSN function~\cite{8}
to construct the conditional distribution of the probability of the production 
of the observed value of asymmetry $\hat A$ by the initial asymmetry $A$.
We assume that an integral luminosity is a constant $\mu_1 + \mu_2 = const$.
The parameters $\mu_1$ and $\mu_2$ are chosen in accordance with the given
initial asymmetry $A$.

\begin{figure}[htpb]
  \begin{center}
\includegraphics[width=0.2\textwidth]{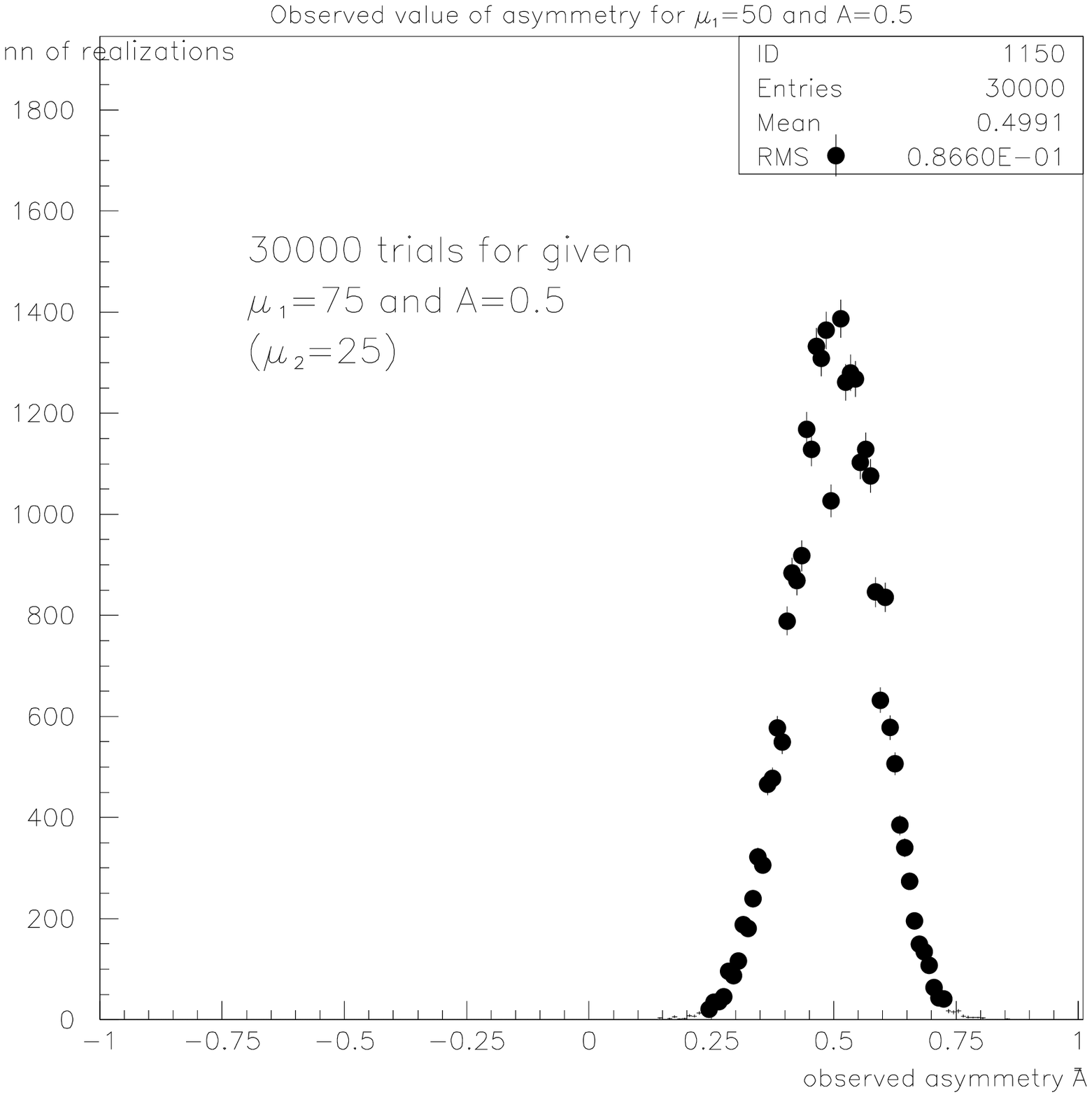} 
\includegraphics[width=0.2\textwidth]{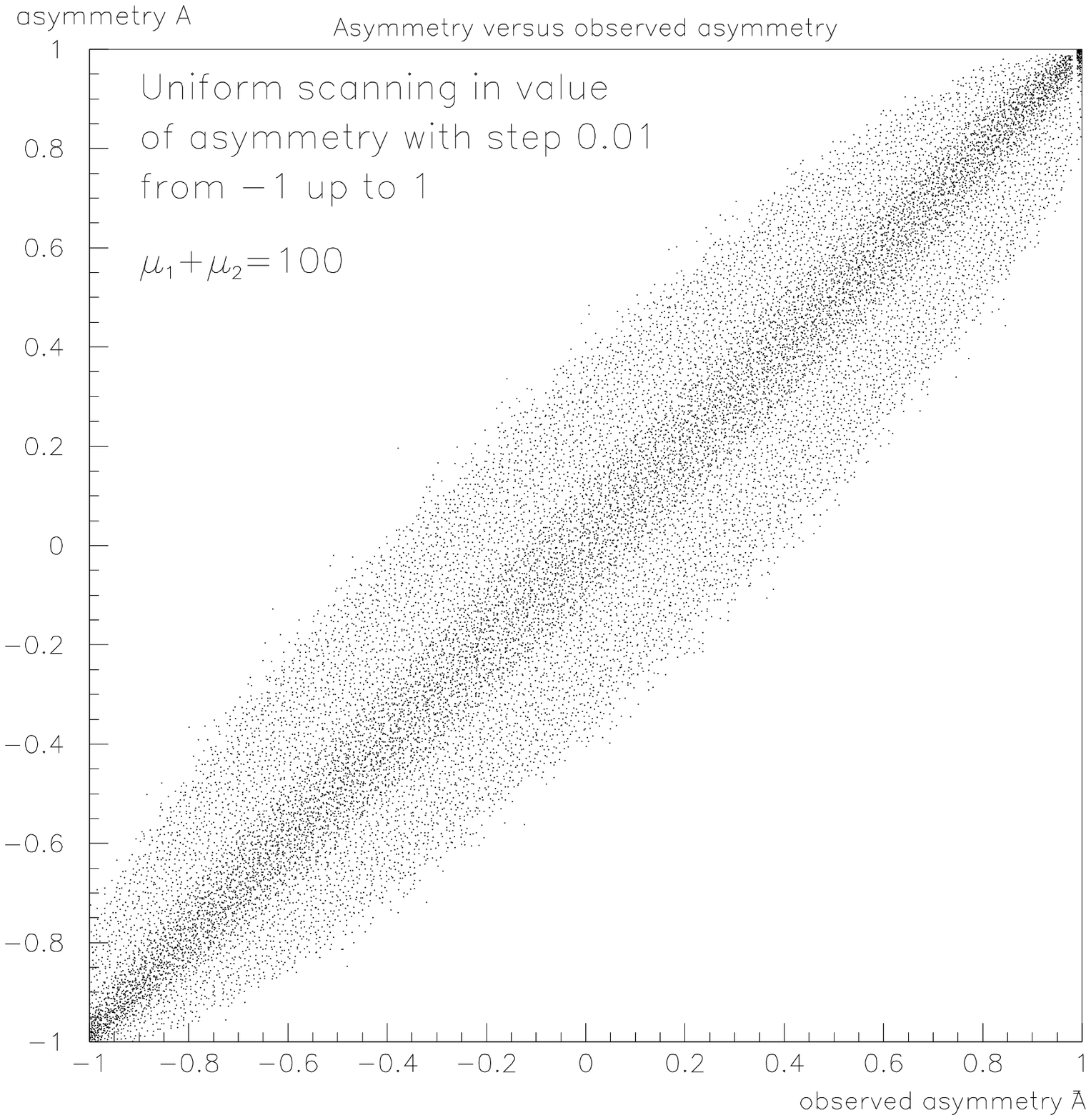} 
\caption{The observed asymmetry $\hat A$ for the case $A=0.5$ 
($\mu_1 + \mu_2 = 100$) (left). The distribution of observed asymmetry 
$\hat A$ versus the initial asymmetry $A$ (right).}
    \label{fig:1} 
  \end{center}
\end{figure}

In Fig.1 (left) the distribution of $\hat A$ for given values 
of $\mu_1 + \mu_2 = 100$ and $A = 0.5$ is shown. 
The distribution of the observed asymmetry $\hat A$ versus the initial 
asymmetry $A$ (Fig.1, right) shows the result of the full scanning.
The distribution of the probability of the initial asymmetries $A$ to 
produce the observed value of $\hat A = 1$ in case of 
$\mu_1 + \mu_2 =10$ is presented in Fig.2 (left).
This figure clearly shows the difference between the most probable value of the
initial asymmetry (A=1) and the mean value of the initial asymmetry (A=0.76).
As seen in Fig.2 (right), the r.m.s. (root-mean-square) 
of the distribution of the initial asymmetry $A$ 
is dependent on the observed value of asymmetry $\hat A$. This 
distribution characterizes the resolution of the determination of 
$A$ by the observed value $\hat A$.
The dependence of the initial asymmetry $A$ on the observed asymmetry
$\hat A$ for $\mu_1 + \mu_2 = 60$ can be seen in Fig.3. The deviation 
from the straight line is essentially dependent on the integral luminosity. 

\begin{figure}[htpb]
  \begin{center}
\includegraphics[width=0.2\textwidth]{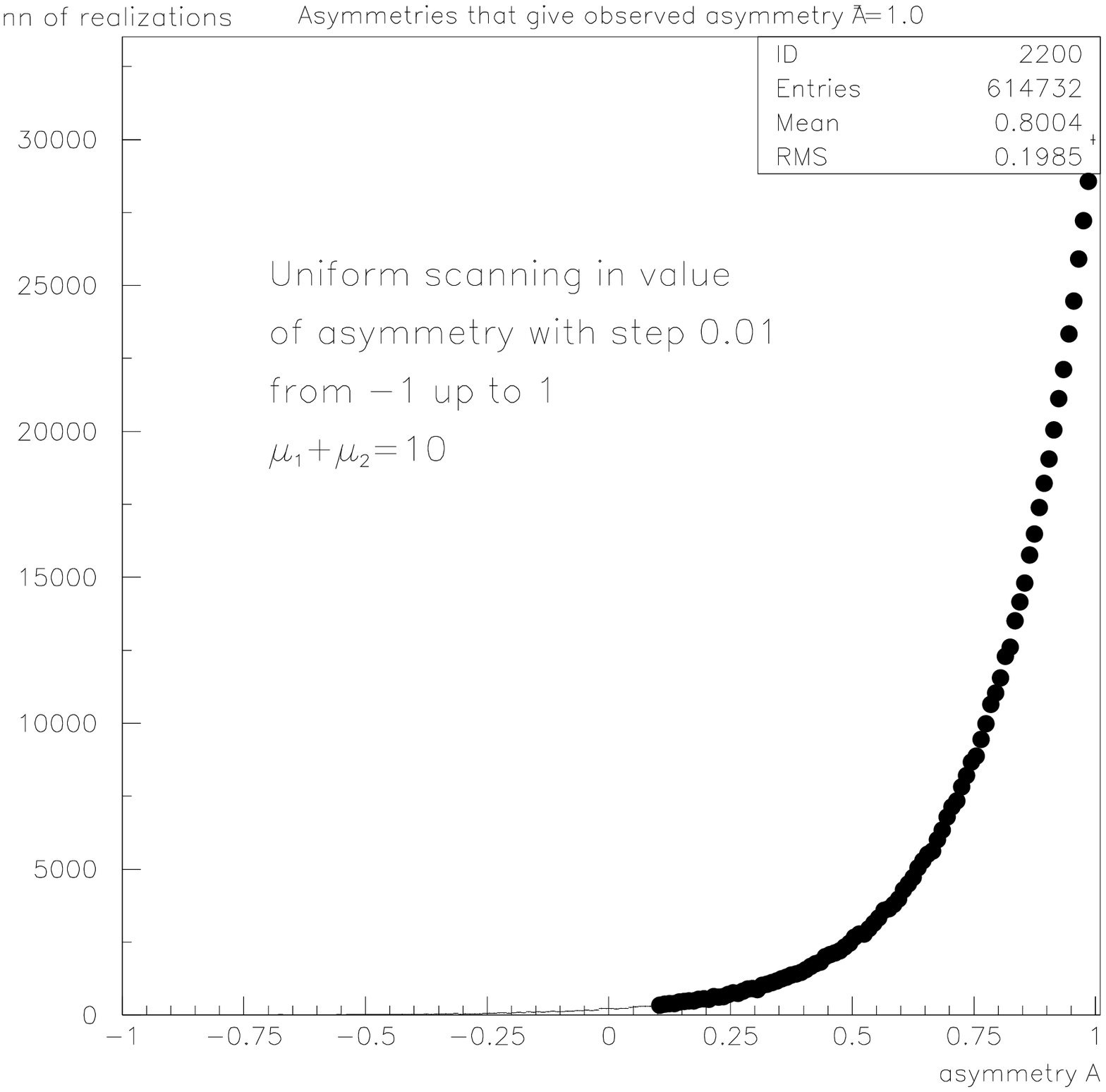} 
\includegraphics[width=0.2\textwidth]{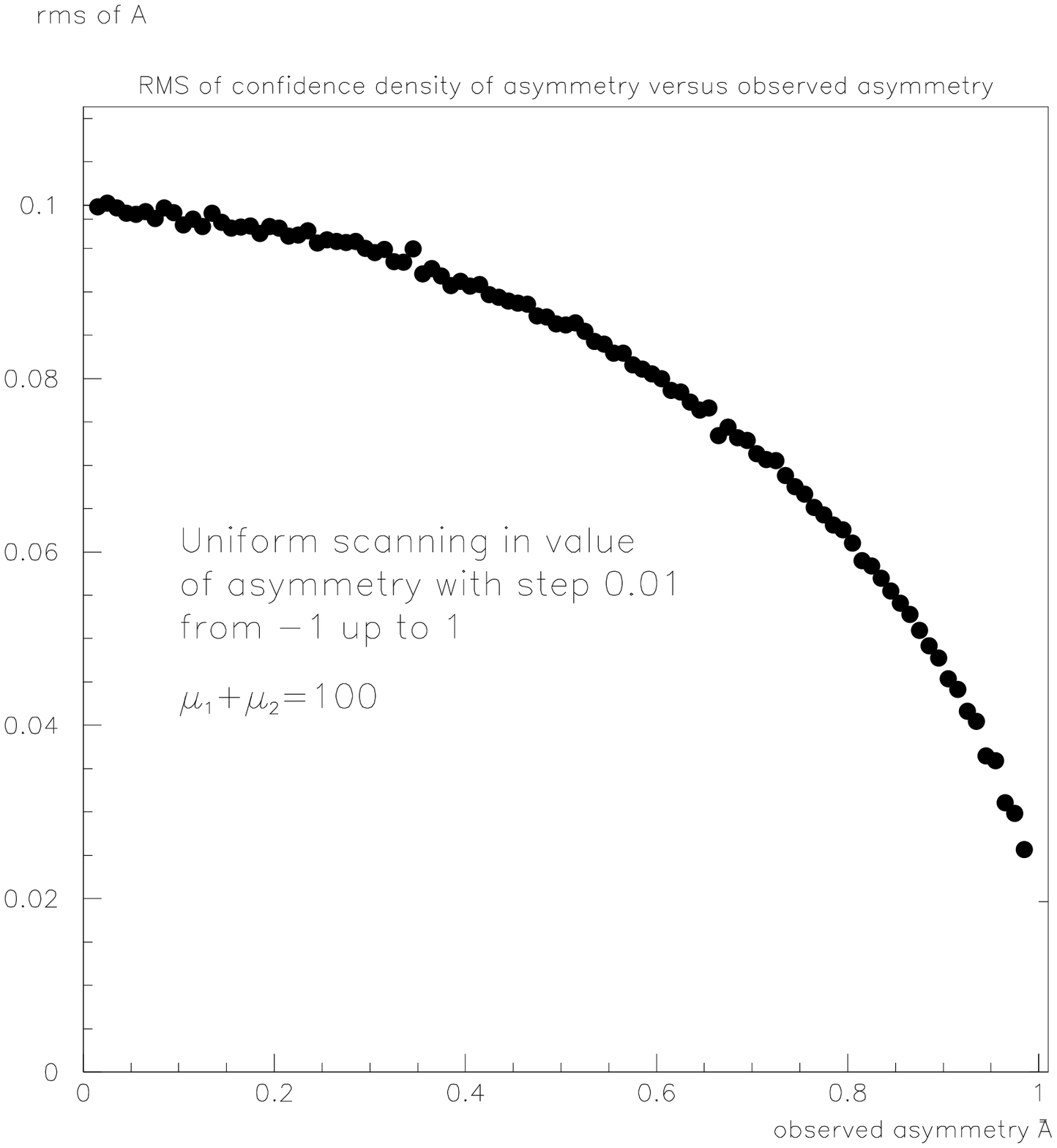} 
\caption{The initial asymmetry $A$ for the case $\hat A=1$, i.e.
$n_2=0$ for $\mu_1 + \mu_2 = 10$ (left). The r.m.s of confidence density 
versus observed asymmetry for $\mu_1 + \mu_2 = 100$ (right).}
    \label{fig:2} 
  \end{center}
\end{figure}

\begin{figure}[htpb]
  \begin{center}
           \resizebox{4.0cm}{!}{\includegraphics{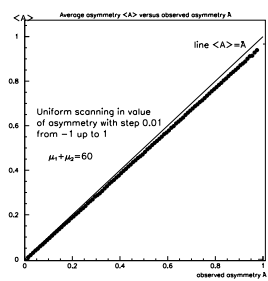}} 
\caption{
The dependence of initial asymmetry 
$A$ on the observed asymmetry $\hat A$  ($\mu_1 + \mu_2 = 60$).}
    \label{fig:3} 
  \end{center}
\end{figure}

\section{Conclusions}

The Monte Carlo experiment confirms the presence of the bias between the mean 
value of the initial asymmetry and the observed asymmetry.
The conditional distribution of the probability of the initial asymmetry $A$ 
to give the observed value $\hat A$ has an asymmetric shape for 
large values of $\hat A$. 
The resolution of the determination of the initial asymmetry $A$
by the observed value $\hat A$ is 
dependent on the value of the observed asymmetry. 
We propose a simple formula Eq.~(\ref{eq:3})
for correction of the observed 
asymmetry. The correct account for the uncertainty of the observed
value must use the distribution of the initial asymmetry, i.e. the 
reconstructed confidence density of the parameter $A$ (see, Fig.2, left).

\section*{Acknowledgments}   

The authors are grateful to V.A.~Kachanov, V.A.~Matveev and 
V.F.~Obraztsov for the interest and useful comments, S.S.~Bityukov, 
J. Cobb, S.V.~Erin, Yu.M. Kharlov, V.A.~Taperechkina, M.N.~Ukhanov 
for fruitful discussions and Jennifer Roper for help in preparing
the paper. This work has been particularly supported 
by grants RFBR 04-01-97227 and RFBR 04-02-16381-a. 

\balance

\end{document}